\documentclass[]{aa}
\usepackage{epsfig}
\begin{document}
\def\gsim{\vcenter{\hbox{$>$}\offinterlineskip\hbox{$\sim$}}}
\thesaurus{06(08.03.1; 08.03.4; 08.13.2; 08.16.4; 11.13.1; 13.09.6)}
\title{IRAS04496$-$6958: A luminous carbon star with silicate dust in the
       Large Magellanic Cloud\thanks{This paper is based on observations with
       the Infrared Space Observatory (ISO). ISO is an ESA project with
       instruments funded by ESA member states (especially the PI countries:
       France, Germany, The Netherlands and the United Kingdom) and with the
       participation of ISAS and NASA.}}
\author{Norman R. Trams\inst{1}, Jacco Th. van Loon\inst{2}, Albert A.
        Zijlstra\inst{3}, Cecile Loup\inst{4}, M.A.T. Groenewegen\inst{5},
        L.B.F.M. Waters\inst{2,6}, Patricia A. Whitelock\inst{7}, J.A.D.L.
        Blommaert\inst{8}, Ralf Siebenmorgen\inst{8} \and Astrid
        Heske\inst{8}}
\institute{Integral Science Operations Centre, Astrophysics Div., Science
           Dep., ESTEC, P.O.Box 299, NL-2200 AG Noordwijk, The Netherlands
      \and Astronomical Institute, University of Amsterdam, Kruislaan 403,
           NL-1098 SJ Amsterdam, The Netherlands
      \and University of Manchester Institute of Science and Technology,
           P.O.Box 88, Manchester M60 1QD, United Kingdom
      \and Institut d'Astrophysique de Paris, 98bis Boulevard Arago, F-75014
           Paris, France
      \and Max-Planck Institut f\"{u}r Astrophysik, Karl-Schwarzschild
           Stra{\ss}e 1, D-85740 Garching bei M\"{u}nchen, Germany
      \and Space Research Organization Netherlands, Landleven 12, NL-9700 AV
           Groningen, The Netherlands
      \and South African Astronomical Observatory, P.O.Box 9, 7935
           Observatory, South Africa
      \and ISO Data Centre, Astrophysics Division, Science Department of ESA,
           Villafranca del Castillo, P.O.Box 50727, E-28080 Madrid, Spain}
\offprints{N.R.\ Trams}
\date{Received date; Accepted date}
\maketitle
\markboth{Trams et al.: Luminous carbon star with silicate dust in LMC}
         {Trams et al.: Luminous carbon star with silicate dust in LMC}
\begin{abstract}

We describe ISO observations of the obscured Asymptotic Giant Branch (AGB)
star IRAS04496$-$6958 in the Large Magellanic Cloud (LMC). This star has been
classified as a carbon star. Our new ISOCAM CVF spectra show that it is the
first carbon star with silicate dust known outside of the Milky Way. The
existence of this object, and the fact that it is one of the highest
luminosity AGB stars in the LMC, provide important information for theoretical
models of AGB evolution and understanding the origin of silicate carbon stars.

\end{abstract}

\section{Introduction}

Asymptotic Giant Branch (AGB) carbon stars are produced following $3^{rd}$
dredge-up in thermally pulsing stars (e.g.\ Iben \& Renzini 1983). The star
changes from oxygen- to carbon-rich when sufficient carbon has been mixed-in
with the stellar mantle to yield an abundance ratio C/O$>1$. The chemistry of
the dust in the circumstellar envelope (CSE) changes accordingly. The change
occurs at smaller core-mass --- or lower luminosity --- for lower metallicity
stars. Clear evidence for this comes from observations of clusters in the
Large Magellanic Cloud (LMC) that contain both carbon and M-type stars (Lloyd
Evans 1984; Marigo et al.\ 1996).

Surprisingly, silicate emission from oxygen-rich dust was discovered in the
IRAS Low Resolution Spectra of several galactic carbon stars (Little-Marenin
1986; Willems \& de Jong 1986). Willems \& de Jong interpreted these
``silicate carbon stars'' as direct evidence for a fast transition of M-type
AGB stars into carbon stars, but timescales of decades for the silicate
emission from an expanding detached oxygen-rich CSE to fade away are difficult
to reconcile with the lifetimes of silicate carbon stars (Little-Marenin et
al.\ 1987; Le Bertre et al.\ 1990). Hence the oxygen-rich material must be
stored in a stationary component. Many galactic silicate carbon stars are
$^{13}$C-enhanced, J-type, carbon stars (Lambert et al.\ 1990). Unlike
genuine, N-type, carbon stars that form on the AGB, J-type carbon stars are
thought to have become carbon-enriched as a result of binary evolution. The
presence of a mass-losing oxygen-rich companion star has been ruled out
observationally for a number of galactic silicate carbon stars (Noguchi et
al.\ 1990; Engels \& Leinert 1994). The presently most supported explanation
for the silicate carbon star phenomenon is that of a keplerian disk of
oxygen-rich material, surrounding a binary including a faint companion (Lloyd
Evans 1990). The oxygen-rich dust may originate from mass loss at a time when
the carbon star was still oxygen rich (Lloyd Evans 1990).

The dust-enshrouded AGB star IRAS04496$-$6958 was recently discovered to be a
luminous carbon star in the LMC by van Loon et al.\ (1998, 1999) on the basis
of ground-based (CTIO) 3 $\mu$m spectroscopy, after having been selected and
confirmed to be an AGB star by Loup et al.\ (1997) and Zijlstra et al.\
(1996), respectively. The carbon star nature of this object has been confirmed
by Groenewegen \& Blommaert (1998) using optical spectroscopy. At $M_{\rm
bol}=-6.8$ mag it is the brightest known magellanic carbon star and very close
to the maximum AGB luminosity ($M_{\rm bol}\sim-7$ mag). We here present
compelling evidence for the presence of oxygen-rich dust close to this
remarkable carbon star, making it the first known extra-galactic silicate
carbon star.

\begin{table*}
\caption[]{ISO 12, 25 and 60 $\mu$m photometry (in Jy) of IRAS04496$-$6958.
The near-IR magnitudes are deduced from light-curves obtained at SAAO
($JD-2,450,000={\rm orbit}+38$), and are on the SAAO photometric system
(Carter 1990). Values between parentheses are 1-$\sigma$ errors.}
\begin{tabular}{lllllllllll}
\hline\hline
 $JD$                          &
 $J [mag]$                     &
 $H [mag]$                     &
 $K [mag]$                     &
 $L [mag]$                     &
 $F_{12}$(CAM)                 &
 $F_{25}$(PHOT)                &
 $F_{60}$(chop)                &
 $F_{60}$(map)                 &
 Spectrum                      \\
\hline
 195                           &
 \llap{1}3.00(0.05)            &
 \llap{1}0.90(0.05)            &
         9.50(0.04)            &
         7.70(0.04)            &
           0.269(0.002)        &
           0.126(0.010)        &
           0.252(0.154)        &
                               &
           PHOT                \\
 605                           &
 \llap{1}2.40(0.05)            &
 \llap{1}0.40(0.05)            &
         8.95(0.05)            &
         7.60(0.05)            &
                               &
                               &
                               &
                               &
         CAM (A)               \\
 905                           &
 \llap{1}2.90(0.10)            &
 \llap{1}1.00(0.10)            &
         9.40(0.05)            &
         7.80(0.05)            &
                               &
                               &
                               &
         0.223(0.123)          &
         CAM (B)               \\
\hline
\end{tabular}
\end{table*}

\section{ISO observations}

The spectral energy distribution (SED) of this object peaks in the infrared.
Therefore in order to properly model the spectrum we obtained photometric and
spectro-photometric observations with the European Infrared Space Observatory
(ISO, see Kessler et al.\ 1996), using the ISOCAM (Cesarsky et al.\ 1996) and
ISOPHOT (Lemke et al.\ 1996) instruments.

The photometric observations at 12 $\mu$m (using ISOCAM filter LW10) and 25
$\mu$m (using ISOPHOT) were obtained on April 22, 1996. The 12 $\mu$m
observation was done using $3^{\prime\prime}$ pixels, and a total on-source
integration time of 50 s split in 25 two-s integration intervals. The 25
$\mu$m ISOPHOT observation was done using the P2 detector, a
$52^{\prime\prime}$ aperture, and triangular chopping with a chopper throw of
$90^{\prime\prime}$. The on-source integration time was 64 s. The 60 $\mu$m
photometry was obtained using the PHOT-C100 detector using two different
methods. One observation (April 22, 1996) was done using chopping mode with
triangular chops and a chopping angle of $150^{\prime\prime}$ and an on-source
integration time of 64 s. Another observation (April 1, 1998) was done with a
$3\times3$ raster map with $46^{\prime\prime}$ raster steps and an integration
time per pointing of 128 s, giving an effective on-source integration time of
$\sim1100$ s.

Spectro-photometric observations of the source were obtained with ISOPHOT-S
and the ISOCAM CVF. The PHOT-S spectrum (April 22, 1996) was done in staring
mode with an on-source integration time of 512 s. Two ISOCAM CVF spectra were
obtained. The first ISOCAM spectrum (June 5, 1997, hereafter ``spectrum A'')
spans the wavelength range from 7 to 14 $\mu$m, the second (April 1, 1998,
hereafter ``spectrum B'') from 5 to 17 $\mu$m. The $6^{\prime\prime}$ pixel
field of view was used, with an integration time per spectral point of 50 and
70 s, respectively.

%
% FIGURE 1
%
\begin{figure}[tb]
\centerline{\psfig{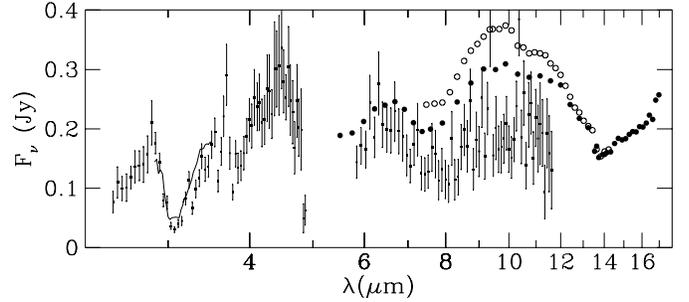}}
\caption[]{IR spectra of IRAS04496$-$6958. The ISOPHOT spectrum is plotted
with errorbars. The ISOCAM spectra A and B are plotted using open and solid
circles, respectively. A 3 $\mu$m spectrum taken at CTIO on December 24, 1996
(van Loon et al.\ 1999) is overplotted as a solid curve, after scaling to the
PHOT-S continuum level.}
\end{figure}

The data was processed using standard processing routines in the PHOT
Interactive Analysis (PIA\footnote{PIA is a joint developement by the ESA
Astrophysics Division and the ISOPHOT consortium led by the Max Planck
Institute for Astronomy (MPIA), Heidelberg. Contributing ISOPHOT Consortium
Institutes are DIAS, RAL, AIP, MPIK and MPIA.}) and CAM Interactive Analysis
(CIA\footnote{CIA is a joint developement by the ESA Astrophysics Division and
the ISOCAM consortium led by the ISOCAM PI, C.\ Cesarsky, Direction des
Sciences de la Materie, C.E.A., France.}) software. The CAM-CVF spectra were
constructed using a $3\times3$ pixel$^2$ software aperture and applying a
correction for the wavelength dependence of the point spread function. We
corrected the PHOT-S spectrum for the background as derived from the CAM-CVF
data, accounting for the annual modulation of the zodiacal light using
COBE/DIRBE weekly all-sky maps (see also Trams et al.\ 1999). The photometric
observations are listed in Table 1. The ISO observations are supplemented with
ground-based J, H, K, and L-band observations made at the South African
Astronomical Observatory (SAAO), interpolated to the same epochs as the
various ISO observations.

The spectra are presented in Fig.\ 1. Also plotted is a spectrum around 3
$\mu$m obtained at CTIO (van Loon et al.\ 1999), after scaling to match the
approximate continuum level in the PHOT-S spectrum. We believe that the PHOT-S
spectrum longward of $\sim6$ $\mu$m has been under-estimated, and possibly
distorted, due to difficulties in determining the stabilised signal at such
low flux density levels.

\section{Discussion}

\subsection{Properties of the circumstellar dust of IRAS04496$-$6958}

%
% FIGURE 2
%
\begin{figure}[tb]
\centerline{\psfig{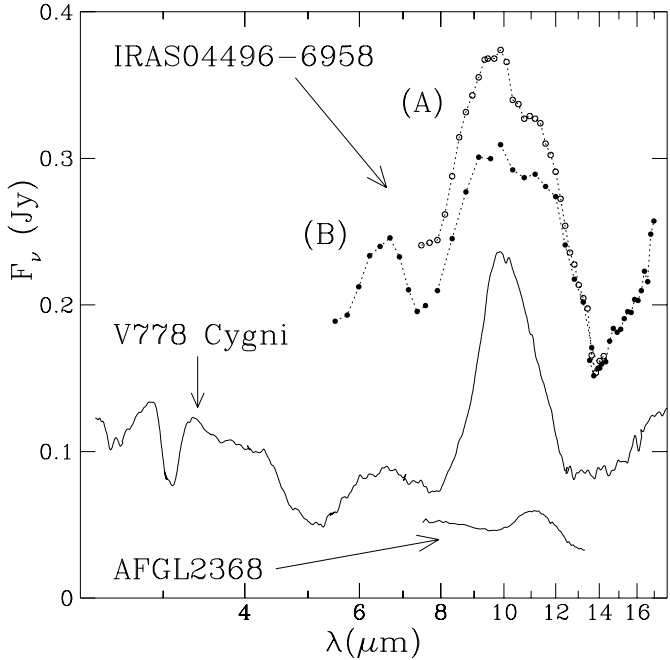}}
\caption[]{The CAM-CVF spectra of IRAS04496$-$6958 compared to the SWS
spectrum ($\div200$) of the silicate carbon star V778 Cyg (Yamamura et al.\
1997) and the UKIRT spectrum ($\div4000$) of the carbon star AFGL2368 (Speck
et al.\ 1997).}
\end{figure}

The PHOT-S and CTIO spectra show the strong 3 $\mu$m feature from HCN and
C$_2$H$_2$ (Fig.\ 1), but the long wavelength part of the PHOT-S spectrum is
rather noisy. The CAM-CVF spectra, however, show a prominent emission feature
between 9 and 12 $\mu$m with a small dip near 11 $\mu$m. For comparison we
also plot in Fig.\ 2 the ISO SWS spectrum ($\div200$) of the galactic silicate
carbon star V778 Cyg, taken from Yamamura et al.\ (1997), which shows strong
silicate emission from oxygen-rich dust around 10 $\mu$m. The feature in
IRAS04496$-$6958 extends to longer wavelengths than in V778 Cyg, and closely
resembles that of another galactic silicate carbon star, CS1003 (Hen 83,
IRAS08002$-$3803; see Little-Marenin 1986 and Willems \& de Jong 1986). We
also plot in Fig.\ 2 the ground-based UKIRT spectrum ($\div4000$) of the
galactic carbon star AFGL2368, taken from Speck et al.\ (1997), which shows a
prominent silicon carbide (SiC) emission feature around $\sim11.5 \mu$m that
is common in carbon stars (see e.g.\ Little-Marenin 1986; Yamamura et al.\
1997). The shape of the 9-12 $\mu$m emission feature in IRAS04496$-$6958 may
be explained by assuming that the feature is a composition of the silicate and
SiC features. Alternative explanations include large silicate grains (Forrest
et al.\ 1975; Papoular \& P\'{e}gouri\'{e} 1983), crystalline olivines (Koike
et al.\ 1981) and corundum (AlO) grains (Onaka et al.\ 1989). We note that
similarly shaped emission is observed in the spectra of a wide variety of
objects: the S star RT Sco and MS stars (Little-Marenin \& Little 1988), the
Sil$^+$ and Sil$^{++}$ classes of M-type Mira variables (Little-Marenin \&
Little 1990), $\beta$ Pictoris (Knacke et al.\ 1993; Fajardo-Acosta \& Knacke
1995), inter-planetary dust particles (Sandford 1988) and comet Halley
(Campins \& Ryan 1989).

Absorption against the photosphere by HCN and C$_2$H$_2$ is seen at 3.1, 3.8
and 8 $\mu$m (Fig.\ 1). On the long-wavelength end of the emission feature the
13.7 $\mu$m absorption due to C$_2$H$_2$ is seen, which is commonly observed
in the spectra of galactic carbon stars (Yamamura et al.\ 1997). This
absorption is seen against the dust continuum --- that dominates over the
photospheric continuum at these long wavelengths --- indicating that the
molecules are abundant throughout the dusty CSE. It is absent in V778 Cyg.
This may be understood if the molecules-to-dust ratio is larger at lower
metallicity, possibly because depletion of molecules into dust grains is less
severe at smaller dust-to-gas ratios.

%
% FIGURE 3
%
\begin{figure}[tb]
\centerline{\psfig{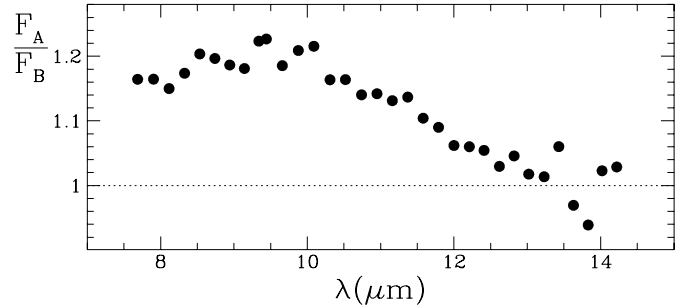}}
\caption[]{The ratio of the two CAM-CVF spectra of IRAS04496$-$6958. The
spectra are variable up to $\sim13$ $\mu$m, where the variability ceases
because stationary dust emission becomes dominant.}
\end{figure}

IRAS04496$-$6958 is a Long Period Variable with a period of $\sim710$ d and a
K-band amplitude of $\sim0.9$ mag (Whitelock et al., in preparation). The two
CVF spectra are taken at different phases in the lightcurve, with spectrum A
closer to maximum light. Their ratio is plotted in Fig.\ 3. The maximum
difference is reached between 9 and 10 $\mu$m, and no difference is seen for
wavelengths $\gsim13$ $\mu$m. This suggests that a significant part of the
variability around 10 $\mu$m is due to a variable emission feature, whilst
beyond 13 $\mu$m non-variable dust continuum emission dominates. The CVF ratio
around 10 $\mu$m is 1.2, which equals the ratio of L-band flux densities at
these two epochs (the K-band flux density ratio is 1.5). This may be compared
to N- (10 $\mu$m) and L-band amplitudes observed in galactic IR-bright carbon
stars, ${\Delta}N-{\Delta}L=-0.48$ mag (standard deviation 0.23) (Le Bertre
1992), and oxygen stars, ${\Delta}N-{\Delta}L=0.13$ mag (standard deviation
0.20) (Le Bertre 1993). The difference results from the fact that in carbon
stars the L-band includes variable HCN+C$_2$H$_2$ absorption, whereas in
oxygen stars the N-band includes variable silicate emission. The
${\Delta}N-{\Delta}L\sim0$ mag of IRAS04496$-$6958 suggests a contribution of
silicate emission to the variability in the N-band. The 10 $\mu$m variability
of IRAS04496$-$6958 is additional evidence for the silicate carbon star nature
of this star, and 10 $\mu$m variability might provide a new means for finding
or confirming silicate carbon stars.

All IR colours between 1 and 25 $\mu$m of IRAS04496$-$6958 are similar to
those of carbon stars (van Loon et al.\ 1998; Trams et al.\ 1999), whereas
V778 Cyg has colours (Chen et al.\ 1998) more similar to oxygen-rich stars.
Hence the oxygen-rich dust component represents only a minor fraction of the
total dust mass that is contained in the CSE of IRAS04496$-$6958. Its CSE is
considerably thicker than that of known galactic silicate carbon stars, judged
from its very red near-IR colours (Lloyd Evans 1990; Chan \& Kwok 1991). Hence
the observation that all galactic silicate carbon stars are of J-type (Lambert
et al.\ 1990) may be an observational bias against N-type carbon stars with
massive carbon-rich CSEs for which an optical spectrum to determine the
$^{13}$C/$^{12}$C ratio is relatively difficult to obtain.

\subsection{The origin of the oxygen-rich dust around the carbon star
IRAS04496$-$6958}

IRAS04496$-$6958 is special because with $M_{\rm bol}=-6.8$ mag it is a very
luminous carbon star (van Loon et al.\ 1998, 1999). The popular scenario for
the formation of silicate carbon stars in which a J-type carbon star evolves
from a less massive R-type star (Lambert et al.\ 1990) implies that silicate
carbon stars should not be very luminous, which has some observational support
(Barnbaum et al.\ 1991). Therefore, IRAS04496$-$6958 could be a different
class from the galactic silicate carbon stars.

Nuclear burning at the base of the convective envelope ("Hot Bottom Burning",
HBB) reduces the carbon abundance of the mantle by cycling it into nitrogen
(Iben 1981; Iben \& Renzini 1983; Wood et al.\ 1983). Theoretical models show
that it occurs for the most massive AGB stars (Bl\"{o}cker \& Sch\"{o}nberner
1991). Boothroyd et al.\ (1993) predict that HBB prevents the occurence of
carbon stars above M$_{\rm bol}=-6.4$ mag, consistent with the observed
absence of optically bright carbon stars more luminous than M$_{\rm
bol}\sim-6$ mag (Iben 1981; Costa \& Frogel 1996). The existence of luminous
dust-enshrouded carbon stars (van Loon et al.\ 1999) like IRAS04496$-$6958 in
the LMC and IRAS00350$-$7436 in the SMC ($M_{\rm bol}=-6.6$ mag; Whitelock et
al.\ 1989) is explained by mass loss reducing the stellar mantle to below a
critical mass required for the pressure and temperature at the lower
convective boundary to be sufficiently high to support HBB (Boothroyd \&
Sackmann 1992). If such a star experiences another thermal pulse and
accompanying dredge-up of carbon to its surface, it may become a carbon star,
after all (Frost et al.\ 1998; Marigo et al.\ 1998). Hence, IRAS04496$-$6958
has been an oxygen-rich star not longer than a thermal pulse interval of
$\sim10^4$ yr ago (see Vassiliadis \& Wood 1993). This does not exclude the
possibility that emission from the oxygen-rich dust is still observable around
the recently formed carbon star, but the massive carbon-rich CSE around
IRAS04496$-$6958 suggests a relatively long lapse of time since the mass loss
was oxygen rich. Although the ISOPHOT 60 $\mu$m photometry is rather
inaccurate, the high 60 $\mu$m flux density of IRAS04496$-$6958 suggest that
its mass-loss rate was higher in the past, some $10^{3-4}$ yr ago, which would
be consistent with an episode of increased mass loss during a thermal pulse
followed by a considerable period of mass loss at a more moderate rate.

Hence it remains to be seen whether the silicate carbon star nature of
IRAS04496$-$6958 requires a companion star to have captured the oxygen-rich
material in a circumbinary disk, or whether it resulted from single star
evolution of a massive AGB star.

\acknowledgements{This research was partly supported by NWO under Pionier
Grant 600-78-333. We would like to thank Prof.\ Michael Feast for his
contribution to the near-IR observations, Dr.\ Angela Speck for kindly
providing the UKIRT spectrum of AFGL2368 in electronic form, Joana Oliveira
for reading the manuscript, and the referee Dr.\ Takashi Onaka for his
interesting comments that helped improve the paper. (JvL: uma beijoca para o
anjinho)}


\begin{thebibliography}{}
\bibitem[1991]{}
Barnbaum C., Kastner J.H., Morris M., Likkel L., 1991, A\&A 251, 79
\bibitem[1991]{}
Bl\"{o}cker T., Sch\"{o}nberner D., 1991, A\&A 244, L43
\bibitem[1992]{}
Boothroyd A.I., Sackmann I.-J., 1992, ApJ 393, L21
\bibitem[1993]{}
Boothroyd A.I., Sackmann I.-J., Ahern S.C., 1993, ApJ 416, 762
\bibitem[1989]{}
Campins H., Ryan E., 1989, ApJ 341, 1059
\bibitem[1990]{}
Carter B.S., 1990, MNRAS 242, 1
\bibitem[1996]{}
Cesarsky C.J., Abergel A., Agn\`{e}se P., et al., 1996, A\&A 315, L32
\bibitem[1991]{}
Chan S.J., Kwok S., 1991, ApJ 383, 837
\bibitem[1998]{}
Chen P.-S., Xiong G.-Z., Wang X.-H., 1998, Acta Astron. Sin. 39, 202
\bibitem[1996]{}
Costa E., Frogel J.A., 1996, AJ 112, 2607
\bibitem[1994]{}
Engels D., Leinert Ch., 1994, A\&A 282, 858
\bibitem[1995]{}
Fajardo-Acosta S.B., Knacke R.F., 1995, A\&A 295, 767
\bibitem[1975]{}
Forrest W.J., Gillett F.C., Stein W.A., 1975, ApJ 195, 423
\bibitem[1998]{}
Frost C.A., Cannon R.C., Lattanzio J.C., Wood P.R., Forestini M., 1998, A\&A
332, L17
\bibitem[1998]{}
Groenewegen M.A.T., Blommaert J.A.D.L., 1998, A\&A 332, 25
\bibitem[1981]{}
Iben I., 1981, ApJ 246, 278
\bibitem[1983]{}
Iben I., Renzini A., 1983, ARA\&A 21, 271
\bibitem[1996]{}
Kessler M.F., Steinz J.A., Anderegg M.E., et al., 1996, A\&A 315, L27
\bibitem[1993]{}
Knacke R.F., Fajardo-Acosta S.B., Telesco C.M., et al., 1993, ApJ 418, 440
\bibitem[1981]{}
Koike C., Hasegawa H., Asada N., Hattori T., 1981, Ap\&SS 79, 77
\bibitem[1990]{}
Lambert D.L., Hinkle K.H., Smith V.V., 1990, AJ 99, 1612
\bibitem[1992]{}
Le Bertre T., 1992, A\&AS 94, 377
\bibitem[1993]{}
Le Bertre T., 1993, A\&AS 97, 729
\bibitem[1990]{}
Le Bertre T., Deguchi S., Nakada Y., 1990, A\&A 235, L5
\bibitem[1996]{}
Lemke D., Klaas U., Abolins J., et al., 1996, A\&A 315, L64
\bibitem[1986]{}
Little-Marenin I.R., 1986, ApJ 307, L15
\bibitem[1988]{}
Little-Marenin I.R., Little S.J., 1988, ApJ 333, 305
\bibitem[1990]{}
Little-Marenin I.R., Little S.J., 1990, AJ 99, 1173
\bibitem[1987]{}
Little-Marenin I.R., Benson P.J., Dickinson D.F., 1987, ApJ 330, 828
\bibitem[1984]{}
Lloyd Evans T., 1984, MNRAS 208, 447
\bibitem[1990]{}
Lloyd Evans T., 1990, MNRAS 243, 336
\bibitem[1997]{}
Loup C., Zijlstra A.A., Waters L.B.F.M., Groenewegen M.A.T., 1997, A\&AS 125,
419
\bibitem[1996]{}
Marigo P., Girardi L., Chiosi C., 1996, A\&A 316, L1
\bibitem[1998]{}
Marigo P., Bressan A., Chiosi C., 1998, A\&A 331, 564
\bibitem[1990]{}
Noguchi K., Murakami H., Matsuo H., et al., 1990, PASJ 42, 441
\bibitem[1989]{}
Onaka T., de Jong T., Willems F.J., 1989, A\&A 218, 169
\bibitem[1983]{}
Papoular R., P\'{e}gouri\'{e} B., 1983, A\&A 128, 335
\bibitem[1988]{}
Sandford S.A., 1988, in: Dust in the Universe, eds.\ M.E. Bailey \& D.A.
Williams, Cambridge University Press, p193
\bibitem[1997]{}
Speck A.K., Barlow M.J., Skinner C.J., 1997, MNRAS 288, 431
\bibitem[1999]{}
Trams N.R., van Loon J.Th., Waters L.B.F.M., et al., 1999, submitted to A\&A
\bibitem[1998]{}
van Loon J.Th., Zijlstra A.A., Whitelock P.A.W., et al., 1998, A\&A 329, 169
\bibitem[1999]{}
van Loon J.Th., Zijlstra A.A., Groenewegen M.A.T., 1999, A\&A in press
\bibitem[1993]{}
Vassiliadis E., Wood P.R., 1993, ApJ 413, 641
\bibitem[1989]{}
Whitelock P.A., Feast M.W., Menzies J.W., Catchpole R.M., 1989, MNRAS 238, 769
\bibitem[1986]{}
Willems F.J., de Jong T., 1986, ApJ 309, L39
\bibitem[1983]{}
Wood P.R., Bessell M.S., Fox M.W., 1983, ApJ 272, 99
\bibitem[1997]{}
Yamamura I., de Jong T., Justtanont K., Cami J., Waters L.B.F.M., 1997, in:
First ISO Workshop on Analytical Spectroscopy, eds.\ A.M. Heras, K.J. Leech,
N.R. Trams \& M. Perry, ESA SP-419, p313
\bibitem[1996]{}
Zijlstra A.A., Loup C., Waters L.B.F.M., et al., 1996, MNRAS 279, 32
\end{thebibliography}
\end{document}